\documentclass[letterpaper]{article} 
\usepackage{aaai25}  
\usepackage{times}  
\usepackage{helvet}  
\usepackage{courier}  
\usepackage[hyphens]{url}  
\usepackage{graphicx} 
\usepackage{natbib}  
\usepackage{caption} 
\usepackage{algorithm}
\usepackage{algorithmic}
\usepackage{latexsym}
\usepackage{amsmath}
\usepackage{booktabs}
\usepackage{multirow}
\usepackage{multicol}
\usepackage{makecell}
\usepackage{graphicx}
\usepackage{tabularx}
\usepackage{caption}
\usepackage{adjustbox}
\usepackage{newfloat}
\usepackage{listings}
\DeclareCaptionStyle{ruled}{labelfont=normalfont,labelsep=colon,strut=off} 
\floatstyle{ruled}
\newfloat{listing}{tb}{lst}{}
\floatname{listing}{Listing}
\title{CSSinger: End-to-End Chunkwise Streaming Singing Voice Synthesis System Based on Conditional Variational Autoencoder}
\author{
    Jianwei Cui\textsuperscript{\rm 1}\thanks{This work was done at Tencent AI Lab as an internship},
    Yu Gu\textsuperscript{\rm 2},
    Shihao Chen\textsuperscript{\rm 1},
    Jie Zhang\textsuperscript{\rm 1},
    Liping Chen\textsuperscript{\rm 1},
    Lirong Dai\textsuperscript{\rm 1}\thanks{Corresponding author}
}
\affiliations{
    \textsuperscript{\rm 1}NERC-SLIP, University of Science and Technology of China (USTC), Hefei, China\\
    \textsuperscript{\rm 2}Tencent AI Lab\\
    \{jwcui,shchen16\}@mail.ustc.edu.cn,
    colinygu@tencent.com,
    \{jzhang6,lrdai\}@ustc.edu.cn
    
}

\usepackage{bibentry}

\begin{document}

\maketitle

\begin{abstract}
Singing Voice Synthesis (SVS) {aims} to generate singing voices {of high} fidelity and expressiveness. {Conventional SVS systems usually utilize} an acoustic model to transform a music score into acoustic features, {followed by a vocoder to reconstruct the} singing voice. It was recently shown that end-to-end modeling is effective in the fields of SVS and Text to Speech (TTS). In this work, we thus present a fully end-to-end SVS method together with a chunkwise streaming inference to address the latency issue for practical usages. Note that this is the first attempt to fully implement end-to-end streaming audio synthesis using latent representations in VAE. We have made specific improvements to enhance the performance of streaming SVS using latent representations. Experimental results demonstrate that the proposed method achieves synthesized audio with high expressiveness and pitch accuracy in both streaming SVS and TTS tasks. Synthesized audio samples are available at: \url{https://sounddemos.github.io/cssinger}.
\end{abstract}

%
\vspace{-0.5cm}
\section{Introduction}

Singing Voice Synthesis (SVS) refers to the task of synthesizing high-quality singing voices by taking {the music score in terms of, e.g., lyrics, pitch and duration, as input} \cite{9362104,nishimura2016singing,hono2021sinsy,lu2020xiaoicesing,hono2019singing}. Similar{ly to Text to Speech (TTS), the SVS system} typically consists of two components: an acoustic model and a vocoder. The former is responsible for modeling the music score as acoustic features, while the vocoder reconstructs the audio waveform using these acoustic features. In this context, Mel-spectrogram is the commonly-employed acoustic representation.
\begin{figure*}[t]
    \vspace{-1cm}
    \centering
    \includegraphics[width=\textwidth, keepaspectratio]{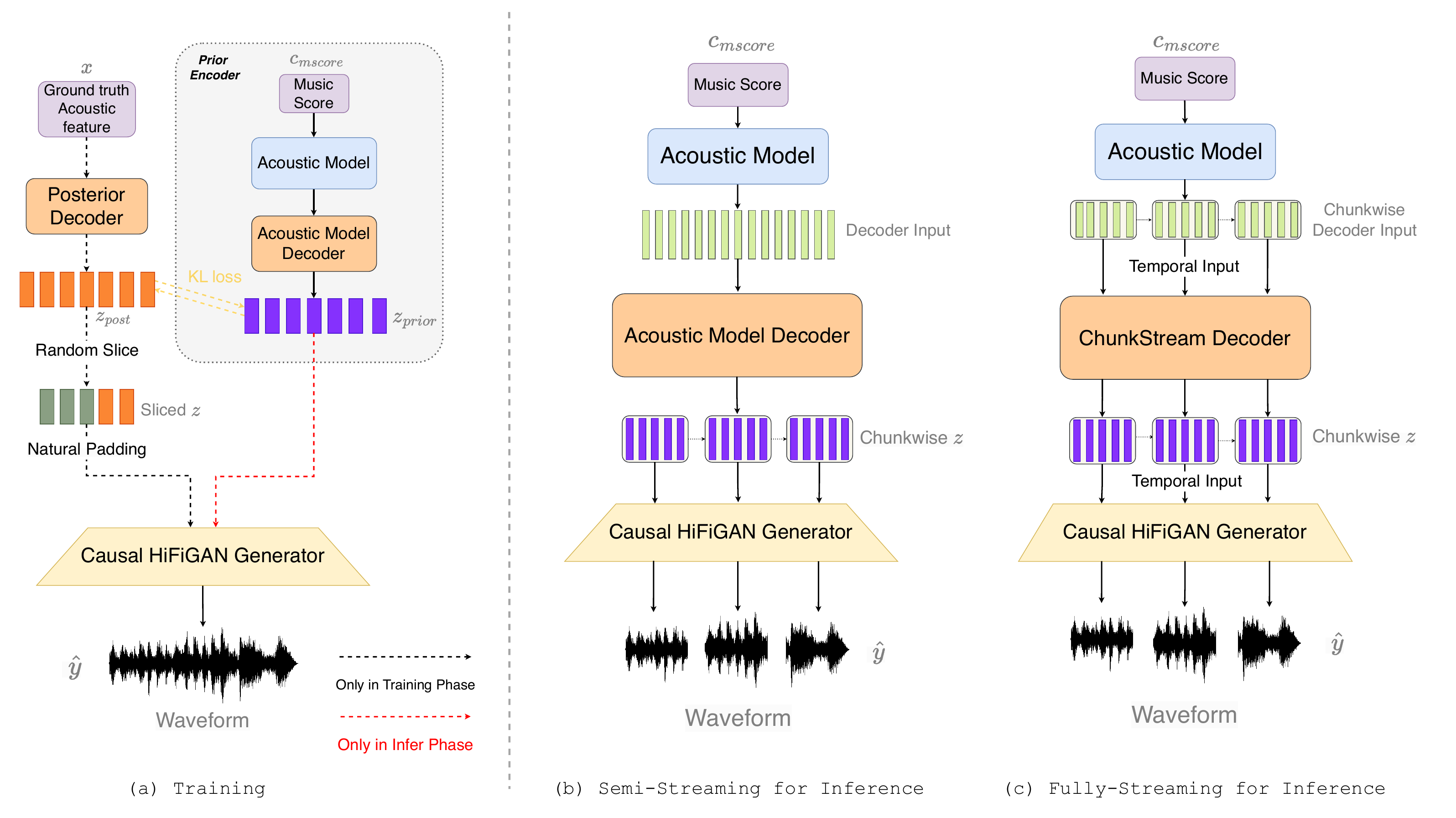}
    \caption{Visualization of the CSSinger Model Structure: (a) Overall Training Process, (b) Semi-Streaming inference pipeline, (c) Fully Streaming inference pipeline.}  
    \label{fig1:archtecture}
    \vspace{-0.5cm}
\end{figure*}

Recently, conditional variational autoencoder (VAE) \cite{ma-etal-2019-flowseq,kingma2013auto} based fully end-to-end systems have been proposed for TTS tasks. For example, unlike two-stage systems that independently develop modules for each stage, VITS \cite{kim2021conditional} employs a prior encoder and a posterior encoder to learn latent space representations and directly generate synthetic audio. The entire system is optimized within a unified training phase. It was shown that the generated synthesized audio by VITS is highly natural.

Based on VITS, VISinger \cite{zhang2022visinger} and VISinger2 \cite{zhang23e_interspeech} introduce an end-to-end SVS system with variational inference. By incorporating a duration predictor and a frame prior network, frame-level means and variances are calculated as acoustic features, facilitating the modeling of rich acoustic variations in singing and achieving natural vocal performance. SiFiSinger \cite{cui2024sifisinger}, as an extension of VISinger, improves the pitch control by introducing a source module to generate F0-controlled excitation signals. It decouples acoustic features from F0 information during modeling, in order to avoid error propagation when predicting acoustic features. For waveform generation, SiFiSinger continuously integrates F0-modulated excitation signals to provide multi-scale pitch guidance during audio synthesis. These can potentially improve the pitch accuracy and naturalness of synthesized singing signals.

{The applicability of these high-performing models is still limited in practice in aspects like computational resources and latency requirements, particularly in the case of deployment on edge devices and/or online network services with data transmission in batches. On one hand, these parallel computation models suffer from a significant computational strain when processing long sequences. On the other hand, their real-time performance is not optimal. Under} these practical constraints, there is a growing need for handling audio synthesis tasks in a streaming and autoregressive manner.
In both acoustic modeling and waveform generation stages, some research focuses on streaming autoregressive modeling. For instance, Tacotron2 \cite{shen2018natural} employs Bi-directional Long Short-term Memory (BLSTM) recurrent networks \cite{schuster1997bidirectional,hochreiter1997long} for acoustic modeling, which suffers from inefficient inference and cannot effectively model long-term dependencies. Similarly, WaveNet \cite{oord2016wavenet} introduces an autoregressive approach for waveform generation but encounters the challenge of slow generation speeds. WaveRNN \cite{kalchbrenner2018efficient} uses a smaller architecture and weight sparsification to accelerate the generation process without significantly compromising quality. The WaveRNN is thus more suitable for real-time application, but requires a complex pipeline.

In this paper, we propose an end-to-end chunk-based streaming SVS system called ChunkStreamSinger (CSSinger), which is somewhat based on SiFiSinger \cite{cui2024sifisinger} (an end-to-end SVS system using conditional VAE). Initially, we explored a semi-streaming framework that directly adapts the HiFi-GAN \cite{kong2020hifi} vocoder for streaming. However, generating audio using latent representations from VAE is not straightforward, particularly for singing voice. In the vocoder, we employed causal transposed convolutions to generate audio in a chunk-wise sequence, but the synthesized audio exhibits significant artifacts, leading to severe quality degradation. We experimentally found that using longer real-valued latent representations as padding for causal convolutions during vocoder training rather than traditional constant padding can substantially improve the audio quality.

Building on this, we thus investigate the chunk streaming acoustic model decoder, which directly models frame-level acoustic features in a chunk-wise streaming manner and captures the corresponding means and variances. This enables the entire system fit a streaming paradigm, from the acoustic model decoder to the HiFi-GAN vocoder, thus avoiding the quadratic time complexity and computational cost faced by attention-based decoders with sequence length. The proposed model achieves subjective and objective metrics that surpass or are on par with the parallel baseline systems on two Chinese singing voice datasets and one TTS dataset, and exhibits a latency significantly lower than the parallel systems. The contributions of this paper can be summarized as follows:
\begin{itemize}
    \item {This is the first attempt for streaming SVS within an end-to-end conditional VAE framework. We consider sequential generation across chunks and allow parallel computation within each chunk.
    \item We address the problem of using latent representations as inputs for causal streaming vocoders, which were showed to be inappropriate for streaming applications.
    \item We employed a chunk streaming acoustic model decoder to implement a fully streaming paradigm. The final model achieved subjective evaluations comparable to those of parallel baseline systems} with the lowest latency.
\end{itemize}
\vspace{-0.2cm}
\section{ChunkStreamSinger}
\subsection{Model Architecture}
Figure~\ref{fig1:archtecture} illustrates the overall model of the proposed method, which is still trained upon the conditional VAE framework of SiFiSinger (see the left part) and follows the semi-streaming scheme based on chunks (see the right part).
\paragraph{Prior Encoder}
In the semi-streaming framework, the structure of the prior encoder remains consistent with the configuration in SiFiSinger. The prior encoder takes the music score \(c_{mscore}\) (lyrics, duration, pitch) as input and extends the input sequence to the frame level via a length regulator, resulting in a prior frame-level latent representation \(z_{prior}\). More specifically, during the training process of the acoustic model, we calculate the prediction losses for F0 and Mel-cepstrum (\(mcep\)), denoted by \(L_{F0}\) and \(L_{mcep}\) respectively, as well as the duration loss \(L_{dur}\). This duration loss is designed to update a duration predictor, facilitating the extension of phone-level representations to frame-level lengths. Subsequently, the decoder of the acoustic (AM Decoder) generates a frame-level prior distribution. The F0 and mcep prediction modules and the AM Decoder are all composed of Feed-Forward Transformer (FFT) \cite{ren2019fastspeech} blocks. We denote the loss function of the prior encoder during the training process as \(L_{am}\), given by

\begin{equation}
L_{am} = L_{F0} + L_{mcep} + L_{dur},\label{eq:l_am}
\end{equation}

\paragraph{Posterior Encoder}
The posterior encoder consists of a series of 1-D convolutional layers and LayerNorm layers. The posterior encoder directly takes the real acoustic features (mcep, F0) as conditional inputs (\(c_{acous}\)) and predicts the mean \(\mu_\phi\left(x\right)\) and variance \(\sigma_\phi\left(x\right)\) of the frame-level posterior distribution, from which the posterior distribution \(z_{post}\) is subsequently sampled. Within this conditional VAE training framework, the training objective is to maximize the variational lower bound of the marginal log-likelihood \(\log p_\theta(x \mid c_{mscore})\). This can be viewed as the sum of the reconstruction loss \(L_{recon}\) and the KL divergence, i.e.,




\begin{equation}
\resizebox{0.85\linewidth}{!}{$
\begin{aligned}
L_{kl} &= \log q_\phi\left(z \mid x\right) - \log p_\theta\left(z \mid c_{mscore}\right),\\
z &\sim q_\phi\left(z \mid x\right)=\mathcal{N}\left(z ; \mu_\phi\left(x\right), \sigma_\phi\left(x\right)\right),
\end{aligned}
$}
\end{equation}
where \(p_\theta\left(z \mid c_{mscore}\right)\) is the prior distribution of the latent variable \(z\) conditioned by \(c_{mscore}\), and \(q_\phi\left(z \mid x\right)\) an approximate posterior distribution.

In the full-stream framework of ChunkStreamSinger, we replace all the convolutions in the Posterior Encoder with causal convolutions, making the entire Posterior Encoder causal. Replacing the convolutions in the Posterior Encoder with causal ones would not significantly affect the modeling process of the posterior distribution. However, to some extent, this operation may help the ChunkStream Decoder  in the full-stream framework to effectively capture the dependencies between chunks, thereby better modeling the prior distribution.
\vspace{-0.5cm}
\begin{figure}[h]
  \includegraphics[width=\columnwidth]{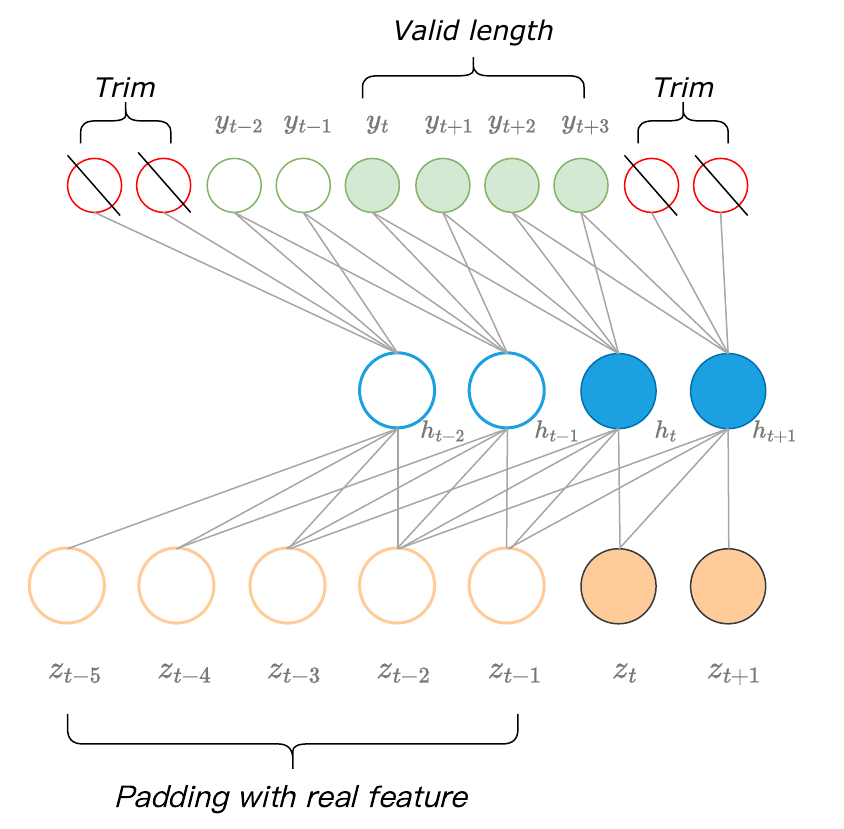}
  \caption{Illustration of the Natural Padding Process}
  \label{fig:naturalpad}
  \vspace{-0.5cm}
\end{figure}

\paragraph{Causal HiFi-GAN Generator}
Based on the public AudioDec\footnote{https://github.com/facebookresearch/AudioDec} \cite{10096509}, we replace both the convolutions and transposed convolutions in the HiFi-GAN generator with causal versions. For the generation of waveforms \(\hat{y}\) in chunks, causal convolutions retain information from previous chunks based on the receptive field size. This  helps to avoid discontinuities in the synthesized audio between chunks. 
The reconstruction loss \(L_{recon}\) is calculated between the mel-spectrograms of the real waveform \(y\) and synthesized waveform \(\hat{y}\) as
\begin{equation}
    L_{recon}  = \|mel(y) - mel(\hat{y})\|_1,
\end{equation}
The training paradigm of adversarial training and the discriminator are consistent with those in SiFiSinger. We compute the generator loss \(L_{adv}(G)\), the discriminator loss \(L_{adv}(D)\) and the feature matching loss \(L_{fm}(G)\). The total loss for training can be expressed as
\begin{equation}
    L=L_{recon}+L_{am}+L_{kl}+L_{adv}(G)+L_{fm}(G),
\end{equation}

\paragraph{Natural Padding}
\label{sec:NP}
In the Semi-Streaming context, we directly segment the latent representation \(z\) into chunks, which are then reconstructed into audio waveforms using the Causal HiFi-GAN generator. Chunking the input significantly reduces the size handled by the generator. Additionally, the entire HiFi-GAN generator is configured with causality during training. This approach can significantly reduce the systematic latency when synthesizing audio without causing discontinuities at the chunk boundaries, thereby preserving the quality of the synthesized audio.

However, we experimentally found that this approach is not straightforward as expected. In case we configure the HiFi-GAN generator with general causal convolutions and causal transposed convolutions to sequentially process chunks of \(z\), the quality of the synthesized audio significantly degrades with extensive artifacts. This is due to the fact that the default causal convolutions typically uses one-sided padding (zero padding or replication padding), where the padding values are usually constant. For mel-spectrograms, a constant value has a specific meaning, e.g., zero represents silence. For latent representations \(z\), a constant value would be ambiguous, as the patterns within \(z\) evolve over training, that is, \(z\) is randomly sliced at each training step. Compared to non-causal convolutions, causal convolutions require more padding on one side. The direct application of common padding strategies would lead  the training process to be more fragile, resulting in a greater mismatch between training and inference. This is the main reason for the degradation of the quality of the synthesized audio.

In the original causal transposed convolution, each layer performs left padding of length \((kernel\_size//stride-1)\) and trims the output sequence on both sides by the length of the \(stride\). This ensures causality and eliminates the impact of padding on the output length. In this work, we introduce a natural padding strategy. Specifically, we eliminate the manual one-side padding operations in all causal convolutions and causal transposed convolutions in the Causal HiFi-GAN generator. In each training step, we extend the beginning of the randomly sliced \(z\) with additional \(z\) values as padding for the input segment of the generator, which passes through several causal convolution and causal transposed convolution layers within the generator. It is unnecessary to precisely calculate the exact receptive field for each layer or the entire generator. As long as the length of the natural padding ensures that the generator's output length exceeds the original length (\(slice\_length \times hop\_size\)), the receptive field overflow does not occur. We then trim the output from the tail to match the original length, in order to produce the final valid output.

\begin{figure}[h]
  \vspace{-0.5cm}
  \includegraphics[width=\columnwidth]{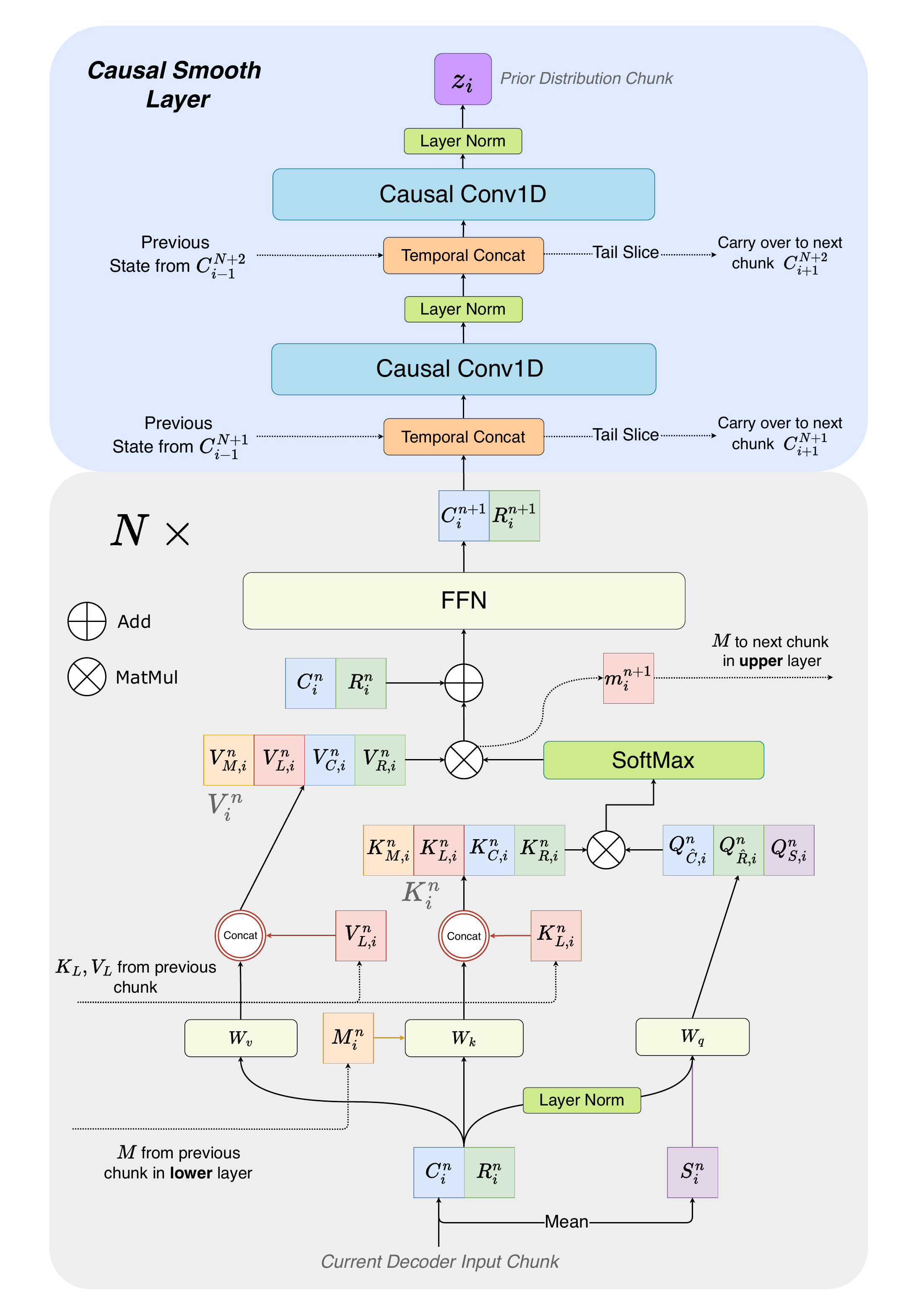}
  \caption{The proposed ChunkStream Decoder, where Current Decoder Input Chunk represents the current input feature vector chunk \(\mathbf{C}_i^n\), \(z_i\)  the corresponding output latent distribution chunk, and \(\mathbf{R}_i^n\)  the right contextual input feature vector extracted from \(\mathbf{C}_i^n\).}
  \label{fig:fullstream}
  \vspace{-0.3cm}
\end{figure}

Figure ~\ref{fig:naturalpad} shows a simple example of the generator processing natural padding. A length-2 input sequence sequentially passes through a causal convolution with \(kernel\_size=4, stride=1\) and a causal transposed convolution with \(kernel\_size=4, stride=2\). Based on the upsampling ratio of the transposed convolution, we expect an valid output length of 4. We extend the beginning of the input sequence with \(z\)-values as natural padding. After obtaining the output, we still trim \(stride\) lengths from both ends and take the valid length of 4 from the tail to get the final output. It can be observed that no receptive field overflow occurs throughout the process, and the valid output is obtained strictly according to the expected upsampling ratio.

\paragraph{Chunkwise Fully-Streaming Framework}
\label{sec:CS decoder}
In the semi-streaming architecture, the AM decoder consists of FFT blocks with an attention \cite{vaswani2017attention} mechanism. The attention mechanism employs a fully parallel computation approach, but the computational complexity is proportional to the square of the sequence length. When handling long sequences, it incurs significant computational resource and time costs. 

Based on the semi-streaming architecture, we further propose a chunk-based fully streaming inference method. As shown in Figure 3, we use the ChunkStream Decoder to transform the generation of the latent representation \(z\) into a chunk-based streaming. Specifically, we draw on the attention mechanism from Emformer \cite{shi2021emformer}, breaking down the complete input feature vectors into several fixed-length chunks. For the calculation of attention, contextual Key, Value and memory bank embeddings extracted from the previous chunk are provided. The memory bank, in an autoregressive manner similar to recurrent neural networks, supplies prior context information. These measures ensure that the attention is performed on a chunk-by-chunk basis and avoid the loss of contextual and global information. The above attention computation process can be summarized as:

\begin{gather}
\left[\hat{\mathbf{C}}_i^n, \hat{\mathbf{R}}_i^n\right] = \operatorname{LayerNorm}\left(\left[\mathbf{C}_i^n, \mathbf{R}_i^n\right]\right), \label{eq:layernorm} \\
\mathbf{K}_i^n = \left[\mathbf{W}_{\mathrm{k}} \mathbf{M}_i^n, \mathbf{K}_{L, i}^n, \mathbf{W}_{\mathrm{k}} \mathbf{C}_i^n, \mathbf{W}_{\mathrm{k}} \mathbf{R}_i^n\right], \label{eq:key} \\
\mathbf{V}_i^n = \left[\mathbf{W}_{\mathrm{v}} \mathbf{M}_i^n, \mathbf{V}_{L, i}^n, \mathbf{W}_{\mathrm{v}} \mathbf{C}_i^n, \mathbf{W}_{\mathrm{v}} \mathbf{R}_i^n\right], \label{eq:value} \\
\mathbf{h}_{\mathrm{C}, i}^n = \operatorname{Attn}\left(\mathbf{W}_{\mathrm{q}} \hat{\mathbf{C}}_i^n, \mathbf{K}_i^n, \mathbf{V}_i^n\right)+\mathbf{C}_i^n, \label{eq:attn_C} \\
\mathbf{h}_{\mathrm{R}, i}^n = \operatorname{Attn}\left(\mathbf{W}_{\mathrm{q}} \hat{\mathbf{R}}_i^n, \mathbf{K}_i^n, \mathbf{V}_i^n\right)+\mathbf{R}_i^n, \label{eq:attn_R} \\
\mathbf{m}_i^{n+1} = \operatorname{Attn}\left(\mathbf{W}_{\mathrm{q}} \mathbf{s}_i^n; \mathbf{K}_i^n, \mathbf{V}_i^n\right), \label{eq:attn_m}\\
\mathbf{C}_i^{n+1}=\operatorname{FFN}(\mathbf{h}_{\mathrm{C} , i}^n),\label{ffn1} \\
\mathbf{R}_i^{n+1}=\operatorname{FFN}(\mathbf{h}_{\mathrm{R}, i}^n),\label{ffn2}
\end{gather}
where \(\mathbf{C}_i^n\) is the $i$-th input chunk of layer \(n\) and \(\mathbf{R}_i^n\) is the right contextual block. The summary vector \(\mathbf{s}_i^n\) is the average of \(\mathbf{C}_i^n\). The memory vector \(\mathbf{M}_i^{n}\) is obtained from the lower layer of the previous chunk. The left contextual key and value matrices \(\mathbf{K}_{L,i}^n\) and \(\mathbf{V}_{L,i}^n\) are directly retrieved from the key and value projections computed in the previous chunk without additional computation. The entire attention operation can be parallelized during training without the need for sequential training.

In this attention mechanism, the contextual dependencies between chunks are captured using key, value and memory bank vectors. To further reduce boundary effects, we introduce the Causal Smooth Layer, which consists of two layers of one-dimensional causal convolutions and LayerNorm layers. The one-dimensional causal convolutions retain features from the same layer of the previous chunk \(\mathbf{C}_{i-1}\) based on the causal receptive field length. Similarly, the tail of the current chunk will also slice features of the causal receptive field length and feed them to the next chunk \(\mathbf{C}_{i+1}\). These measures provide finer smoothing at the boundaries to enhance the naturalness and fluency of the synthesized audio quality.

\vspace{-0.2cm}
\section{Experiments}

In this section, we will present the experimental setup and evaluation results of the proposed method for both SVS and TTS tasks.

\subsection{Dataset}
We use the Opencpop\footnote{https://wenet.org.cn/opencpop/} \cite{wang2022opencpop} dataset, a publicly available high-quality Chinese singing voice dataset designed for SVS systems. It includes 100 Mandarin songs performed by a professional female singer. All songs were phonetically annotated with utterance/note/phoneme boundaries and pitch types. The final dataset comprises 3,756 utterances, totaling 5.2 hours of audio. The test set consists of 5 songs, comprising a total of 206 utterances and official test set partitions. We also use the PopCS \cite{liu2022diffsinger} dataset, a Chinese Mandarin singing voice dataset, which contains 117 songs ($\approx$5.89 hours, all performed by a professional female singer). PopCS provides lyrics and phoneme-level alignments between song pieces, and the corresponding lyrics were obtained using the Montreal Forced Aligner tool (MFA) \cite{mcauliffe2017montreal}. We randomly selected 140 utterances from 10 songs as the test set.

For the TTS task, we use the Baker\footnote{https://en.data-baker.com/datasets/freeDatasets/} dataset, a Chinese female speech dataset ($\approx$12 hours of standard Mandarin female speech), which contains 10,000 sentences and each has 16 characters on average. We select 200 sentences from all the audio recordings to form the test set. The data was collected in a professional recording studio. The dataset provides wav audio files, text annotations and phoneme boundary duration annotation files.

\subsection{Comparison methods}
{We train several systems for comparison:
1) \textbf{Recording}, using the ground-truth singing voice audio;
2) \textbf{SiFiSinger}, a fully parallel inference system used as the baseline \cite{cui2024sifisinger};
3) \textbf{CSSinger-SS}, the semi-streaming method of ChunkStream Singer , where only the HiFi-GAN generator runs in a chunkwise streaming manner;
4) \textbf{CSSinger-SS-NP}, the semi-streaming structure with Natural Padding added to the HiFi-GAN generator;
5) \textbf{CSSinger-FS} (proposed model), the fully streaming framework of ChunkStream Singer, which uses} the ChunkStream Decoder to enable streaming inference in both the generation of latent representation \(z\) and the generator. 

\begin{table}[ht]
\centering
\resizebox{\columnwidth}{!}{%
\begin{tabular}{@{}lccc@{}}
\toprule
\textbf{Models} &
  \begin{tabular}[c]{@{}c@{}}\textbf{Sample} \\ \textbf{Rate}\end{tabular} &
  \begin{tabular}[c]{@{}c@{}}\textbf{MOS}\\ \textbf{(Opencpop)}\end{tabular} &
  \begin{tabular}[c]{@{}c@{}}\textbf{MOS}\\ \textbf{(PopCS)}\end{tabular} \\ \midrule
Recording      & 44.1KHz & 4.220±0.073 & 4.322±0.076 \\ \midrule
SiFiSinger     & 44.1KHz & 3.510±0.097 & 3.436±0.081 \\
CSSinger-SS    & 44.1KHz & 2.415±0.092 & 3.447±0.081 \\
CSSinger-SS-NP & 44.1KHz & 3.165±0.095 & 3.491±0.080 \\
CSSinger-FS    & 44.1KHz & \textbf{3.607±0.091} & \textbf{3.500±0.086} \\ \bottomrule
\end{tabular}%
}
\caption{Subjective MOS tests results with 95\% confidence interval for SVS.}
\label{tab:SVS MOS}
\vspace{-0.5cm}
\end{table}

\subsection{Implementation Details}
For the singing voice datasets Opencpop and PopCS, we used audio sampled at 44.1KHz with 16-bit quantization. For the TTS task, the Baker dataset provides audio sampled at 48KHz, which we resampled to 16KHz. Opencpop offers detailed music scores (lyrics, phoneme, note, duration, slur flag). We followed the same data processing pipeline as SiFiSinger for Opencpop. The PopCS dataset does not provide MIDI information, so we extracted pitch using the phoneme-level timestamps provided and converted these to corresponding note annotations for model input. The Baker dataset provides Chinese text annotations and corresponding pinyin, which we further segmented into initials and finals as phone-level inputs for the model. All three datasets provide phone-level duration information, which can be directly used to train the duration predictor in the model.

The specific configuration of the model is largely consistent with SiFiSinger. The model's hidden size is set to 192, and the channel number in the FFN hidden layer is 768. The Acoustic Model Decoder and ChunkStream Decoder both have 4 attention layers (\(N=4\)). We use diffsptk\footnote{https://github.com/sp-nitech/diffsptk} \cite{sp-nitech2023sptk} to extract 80-dimensional mel-cepstrum (mcep) as the acoustic features. The hop size is set to 512 for the Opencpop and PopCS datasets, while it is set to 256 for the Baker dataset. In the TTS task, there is no note information, so the text encoder only receives phoneme text inputs to predict acoustic features and F0.

{In all experiments, the size of the random slice for training the HiFi-GAN generator is set to 20. In all CSSinger models, the chunk size for both training and inference is also 20. In the fully streaming structure of CSSinger, the left context window size is set to 10 and the right context window size is 4. This also determines the length of the Key and Value vectors (\(\mathbf{K}_{L, i}^n\), \(\mathbf{K}_{R, i}^n\), \(\mathbf{V}_{L, i}^n\), \(\mathbf{V}_{R, i}^n\)) from the left and right contexts (e.g., see Section~\ref{sec:CS decoder}).}

\begin{figure*}[t]
    \vspace{-1cm}
    \centering
    \includegraphics[width=\textwidth, keepaspectratio]{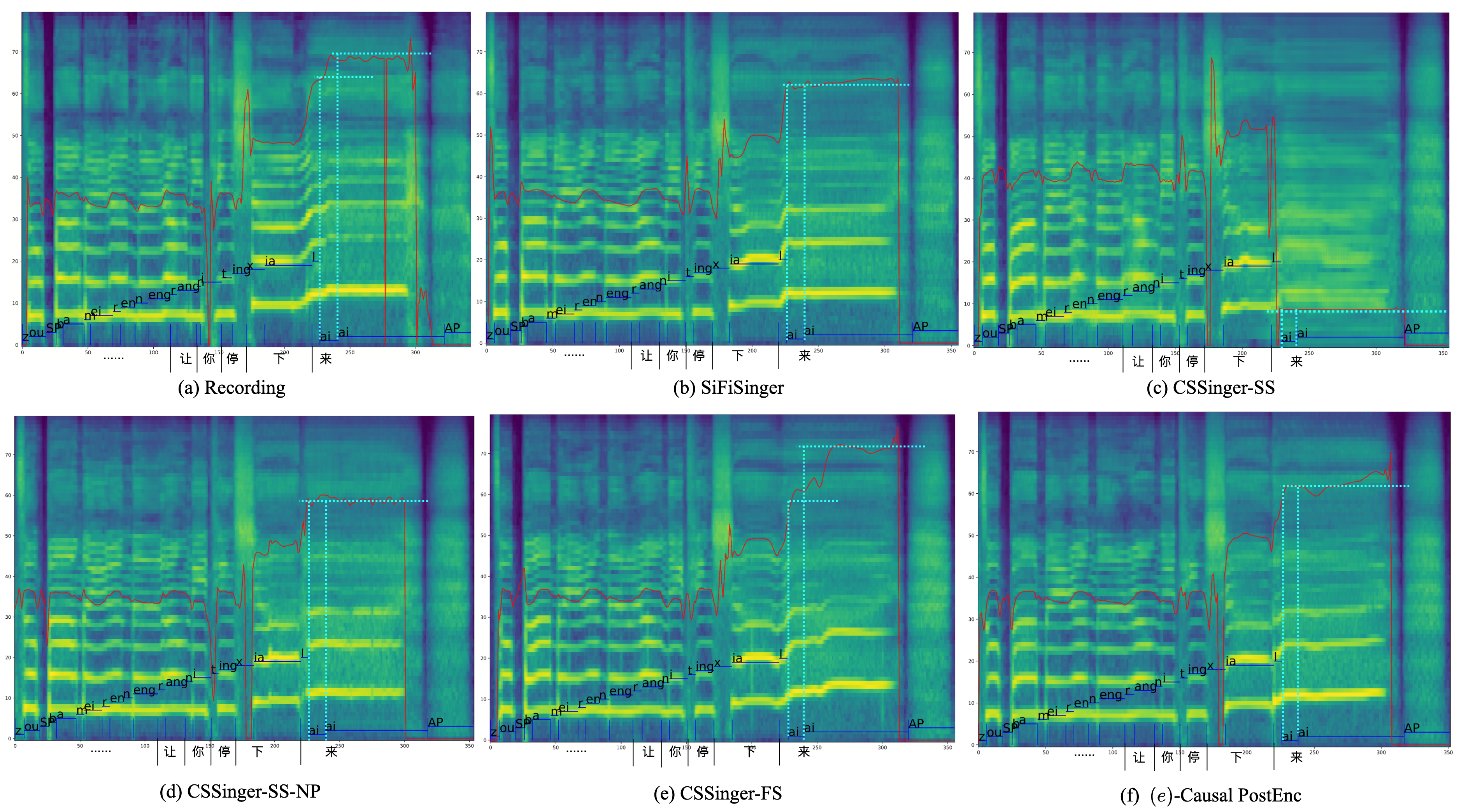}
    \caption{Mel-spectrograms with phoneme boundaries and pitch contour for the same audio sample obtained by comparison models.}  
    \label{fig:melspec}
    \vspace{-0.3cm}
\end{figure*}
\subsection{Main Results \& Analysis}
\paragraph{Evaluation of SVS}
To {verify the efficacy of the proposed model for SVS, we evaluate the mean opinion score (MOS) on the test sets of Opencpop and PopCS. We randomly select 20 samples from each dataset and invite 20 native speaker to perform subjective evaluations. The results are shown in Table~\ref{tab:SVS MOS}, from which it is clear that the  proposed fully streaming structure of ChunkStreamSinger (\textbf{CSSinger-FS}) achieves the best performance on both datasets. It significantly outperforms the semi-streaming structures (\textbf{CSSinger-SS} and \textbf{CSSinger-SS-NP}) and is even better than the fully parallel baseline (\textbf{SiFiSinger}). On the Opencpop dataset, the semi-streaming structure  (\textbf{CSSinger-SS}) suffers from significant degradation in audio quality due to the Causal HiFi-GAN generator that applies one-sided constant padding to the input latent representation \(z\). This causes instability in the
modeling of \(z\) and a significant mismatch between training and inference, resulting in poor performance in MOS tests. As described in Section~\ref{sec:NP}, the introduction of Natural Padding (\textbf{CSSinger-SS}) can partially overcome this, leading to a significant increase in the MOS score. The fully streaming structure also adopts a chunk-streaming approach for modeling \(z\), aligning the behavior paradigm with the process of reconstructing audio waveforms from \(z\). Owing to the ChunkStream Decoder, the latent representation \(z\) is better at capturing local contextual variations and connections, which is thus more useful to produce vivid, natural} and pitch-accurate singing voices.

\begin{table}[ht]

\centering
\resizebox{\columnwidth}{!}{%
\begin{tabular}{@{}llllll@{}}
\toprule
\multicolumn{1}{l}{\multirow{2}{*}{\textbf{Models}}} &
  \multicolumn{1}{c}{\multirow{2}{*}{\textbf{\begin{tabular}[c]{@{}c@{}}F0 \\ RMSE\( \downarrow \)\end{tabular}}}} &
  \multicolumn{1}{c}{\multirow{2}{*}{\textbf{\begin{tabular}[c]{@{}c@{}}F0 \\ Corr\( \uparrow \)\end{tabular}}}} &
  \multicolumn{1}{c}{\multirow{2}{*}{\textbf{\begin{tabular}[c]{@{}c@{}}U/UV \\ Err\( \downarrow \)\end{tabular}}}} &
  \multicolumn{1}{c}{\multirow{2}{*}{\textbf{MSE}\( \downarrow \)}} &
  \multicolumn{1}{c}{\multirow{2}{*}{\textbf{MCD}\( \downarrow \)}} \\
\multicolumn{1}{c}{}         & \multicolumn{1}{c}{} & \multicolumn{1}{c}{} & \multicolumn{1}{c}{} & \multicolumn{1}{c}{} & \multicolumn{1}{c}{} \\ \midrule
SiFiSinger & 34.295               & 0.908                & 0.117                & 1.120                & 6.882                \\
CSSinger-SS         & 56.625               & 0.654                & 0.172                & 1.304                & 8.006                \\
CSSinger-SS-NP      & 29.517               & 0.911                & 0.122                & 1.226                & 7.529                \\
CSSinger-FS         & 28.601               & 0.919                & 0.107                & 1.093                & 6.715                \\ \bottomrule
\end{tabular}%
}
\caption{Objective evaluation on Opencpop dataset}
\label{tab:objective Opencpop}
\vspace{-0.5cm}
\end{table}

We also {conduct objective tests on all test set entries, for which calculate root mean square error (F0 RMSE) and correlation coefficient (F0 Corr) of F0 between the synthesized audio and the ground truth. Besides, we measure the unvoiced/voiced error rate (U/UV Err), the mean square error (MSE) of the Mel-cepstrum and the mel-cepstral distortion (MCD). The results on the two datasets are shown in Tables~\ref{tab:objective Opencpop} and~\ref{tab:objective PopCS}, respectively. It can be seen that on the Opencpop dataset, \textbf{CSSinger-FS} achieves the best performance in all metrics, and the poor audio quality of \textbf{CSSinger-SS} is also reflected in the objective metrics. On the PopCS dataset, the objective metrics of the proposed model also surpasses} the \textbf{SiFiSinger} baseline.

\begin{table}[ht]
\centering
\resizebox{\columnwidth}{!}{%
\begin{tabular}{@{}llllll@{}}
\toprule
\multirow{2}{*}{\textbf{Models}} &
  \multicolumn{1}{c}{\multirow{2}{*}{\textbf{\begin{tabular}[c]{@{}c@{}}F0 \\ RMSE\( \downarrow \)\end{tabular}}}} &
  \multicolumn{1}{c}{\multirow{2}{*}{\textbf{\begin{tabular}[c]{@{}c@{}}F0 \\ corr\( \uparrow \)\end{tabular}}}} &
  \multicolumn{1}{c}{\multirow{2}{*}{\textbf{\begin{tabular}[c]{@{}c@{}}U/UV \\ err\( \downarrow \)\end{tabular}}}} &
  \multicolumn{1}{c}{\multirow{2}{*}{\textbf{MSE}\( \downarrow \)}} &
  \multicolumn{1}{c}{\multirow{2}{*}{\textbf{MCD}\( \downarrow \)}} \\
                        & \multicolumn{1}{c}{} & \multicolumn{1}{c}{} & \multicolumn{1}{c}{} & \multicolumn{1}{c}{} & \multicolumn{1}{c}{} \\ \midrule
SiFiSinger     & 27.064               & 0.872                & 0.144                & 1.430                & 8.783                \\
CSSinger-SS    & 26.593               & 0.876                & 0.144                & 1.426                & 8.756                \\
CSSinger-SS-NP & 25.748               & 0.886                & 0.142                & 1.415                & 8.690                \\
CSSinger-FS    & 26.150               & 0.885                & 0.145                & 1.425                & 8.749                \\ \bottomrule
\end{tabular}%
}
\caption{Objective evaluation of PopCS dataset}
\vspace{-0.3cm}
\label{tab:objective PopCS}

\end{table}

{Figure~\ref{fig:melspec} visualizes the Mel-spectrograms of a synthesized singing voice sample obtained by comparison models, where we include the phoneme boundaries at the frame level and the pitch contour (in red). The last Chinese character in the lyrics consists of the phonemes "\textit{l}" and "\textit{ai}", with "\textit{ai}" occupying two phoneme inputs. The note labels for the two "\textit{ai}" phonemes are "\textit{D\#5/Eb5}" and "\textit{F5}", respectively. This indicates that the ending of the song should be a high pitch transition on the same phoneme "\textit{ai}". From the plot of the recording in Figure~\ref{fig:melspec}(a), we observe the pitch transition of the "\textit{ai}" phoneme with a light blue dashed line on the corresponding pitch contour. Among the tested models, only \textbf{CSSinger-FS} successfully models this upward pitch transition. In contrast, in the synthesized audio obtained by other models, the pitch of the last two "\textit{ai}" phonemes remains unchanged, resulting in a flat pitch ending.

\begin{table*}[ht]
    \vspace{-0.5cm}
    \centering
    \resizebox{\textwidth}{!}{%
    \begin{tabular}{lccccccccc}
        \toprule
        & \multicolumn{3}{c}{\textbf{GPU}} & \multicolumn{3}{c}{\textbf{CPU}} & \multicolumn{3}{c}{\textbf{CPU-Limited}}\\
        \cmidrule(lr){2-4} \cmidrule(lr){5-7} \cmidrule(lr){8-10}
        \textbf{Models} & \makecell{\textbf{Latency}\\\textbf{(s)}\(\downarrow\)} & \makecell{\textbf{Process}\\\textbf{Time(s)}\(\downarrow\)} & \textbf{RTF}\(\downarrow\) & \makecell{\textbf{Latency}\\\textbf{(s)}\(\downarrow\)} & \makecell{\textbf{Process}\\\textbf{Time(s)}\(\downarrow\)} & \textbf{RTF}\(\downarrow\) & \makecell{\textbf{Latency}\\\textbf{(s)}\(\downarrow\)} & \makecell{\textbf{Process}\\\textbf{Time(s)}\(\downarrow\)} & \textbf{RTF}\(\downarrow\) \\
        \midrule
        \textbf{SiFiSinger} & 0.180 & \textbf{0.180} & \textbf{0.038} & 1.508 & \textbf{1.508} & \textbf{0.311} & 4.058 & 4.058 & 0.835 \\
        \textbf{CSSinger-SS} & 0.176 & 0.542 & 0.112 & 0.523 & 1.614 & 0.334 & 1.280 & 3.657 & 0.752 \\
        \textbf{CSSinger-SS-NP} & 0.176 & 0.542 & 0.112 & 0.536 & 1.660 & 0.344 & 1.119 & 3.248 & 0.668 \\
        \textbf{CSSinger-FS} & \textbf{0.051} & 0.476 & 0.099 & \textbf{0.483} & 1.675 & 0.347 & \textbf{0.965} & \textbf{3.082} & \textbf{0.635} \\
        \bottomrule
    \end{tabular}%
    }
    \caption{The inference latency and efficiency indicators with different hardware configurations.}
    \label{tab:latency}
    \vspace{-0.5cm}
\end{table*}
Apart from the pitch accuracy, the proposed model also exhibits a good audio quality in the spectrogram. As Figure~\ref{fig:melspec}(c) shows that \textbf{CSSinger-SS} exhibits extensive artifacts in the harmonic components,  the audio quality might degrade as well as pitch extraction. Due to the introduction of Natural Padding, \textbf{CSSinger-SS-NP} greatly improves the synthesized audio quality. However, due to the inconsistency between the Acoustic Model Decoder and the generator, some discontinuities and spikes appear in the spectrogram. In contrast, the proposed model achieves an audio quality comparable to the ground truth and the baseline.}

\paragraph{Evaluation on Text-to-Speech}
\begin{table}[ht]
\vspace{-0.2cm}
\centering
\small
\begin{tabular}{@{}lcc@{}}
\toprule
\textbf{Models} & \textbf{Sample Rate} & \textbf{MOS} \\ \midrule
Recording      & 16KHz & 4.244±0.075 \\ \midrule
SiFiSinger     & 16KHz & 3.911±0.088 \\
CSSinger-SS    & 16KHz & 3.428±0.079 \\
CSSinger-SS-NP & 16KHz & 3.708±0.082 \\
CSSinger-FS    & 16KHz & 3.828±0.071 \\ \bottomrule
\end{tabular}
\caption{Subjective MOS tests with 95\% confidence interval on the Baker dataset for TTS.}
\label{tab:TTS MOS}
\vspace{-0.3cm}
\end{table}

{To verify the generality of the proposed method, we also conduct experiments on the TTS task. We randomly select twenty samples from the synthesized audio in the Baker test set, and invite twenty native speakers to conduct subjective evaluations. The results in terms of MOS score are shown in Table~\ref{tab:TTS MOS}. It can be seen that the introduction of Natural Padding and the fully streaming framework turns out significant improvements. The corresponding objective evaluations are shown in Table~\ref{tab:objective Baker}. The proposed fully streaming framework achieves the best performance among the CSSinger-related models in objective metrics as well. These  demonstrate the effectiveness of our proposed method for TTS}.

\begin{table}[ht]
\centering
\resizebox{\columnwidth}{!}{%
\begin{tabular}{@{}llllll@{}}
\toprule
\multirow{2}{*}{\textbf{Models}} &
  \multicolumn{1}{c}{\multirow{2}{*}{\textbf{\begin{tabular}[c]{@{}c@{}}F0 \\ RMSE\( \downarrow \)\end{tabular}}}} &
  \multicolumn{1}{c}{\multirow{2}{*}{\textbf{\begin{tabular}[c]{@{}c@{}}F0 \\ Corr\( \uparrow \)\end{tabular}}}} &
  \multicolumn{1}{c}{\multirow{2}{*}{\textbf{\begin{tabular}[c]{@{}c@{}}U/UV \\ err\( \downarrow \)\end{tabular}}}} &
  \multicolumn{1}{c}{\multirow{2}{*}{\textbf{MSE}\( \downarrow \)}} &
  \multicolumn{1}{c}{\multirow{2}{*}{\textbf{MCD}\( \downarrow \)}} \\
                        & \multicolumn{1}{c}{} & \multicolumn{1}{c}{} & \multicolumn{1}{c}{} & \multicolumn{1}{c}{} & \multicolumn{1}{c}{} \\ \midrule
\textbf{SiFiSinger}     & 38.187               & 0.817                & 0.144                & 1.255                & 7.706                \\ 
\textbf{CSSinger-SS}    & 42.687               & 0.736                & 0.199                & 1.481                & 9.099                \\
\textbf{CSSinger-SS-NP} & 42.518               & 0.747                & 0.202                & 1.483                & 9.110                \\
\textbf{CSSinger-FS}    & 41.278               & 0.783                & 0.205                & 1.427                & 8.763
\\ \bottomrule
\end{tabular}%
}
\caption{Objective results on the Baker dataset for TTS.}
\label{tab:objective Baker}
\vspace{-0.5cm}
\end{table}

\paragraph{Latency Evaluation}

{In order to evaluate the latency of the proposed model, we measure the latency from input features to the generation of synthesized audio (Latency in seconds), the total processing time from input features to the completion of audio processing (Process Time) and the Real-Time Factor (RTF) as indicators of systematic efficiency. The GPU environment used in the experiments is a single NVIDIA 32GB V100 GPU, and the CPU is an Intel(R) Xeon(R) Platinum 8255C CPU @2.50GHz with 10 cores and 20 threads. We test the latency metrics in three scenarios: GPU inference (GPU), CPU inference (CPU), and single-core single-thread CPU-limited inference (CPU-Limited). This experiment is conducted on all test set entries of Opencpop, using the same ground-truth duration during inference. 

The latency results are shown in Table~\ref{tab:latency}, from which we see that in the GPU inference scenario, the semi-streaming models (\textbf{CSSinger-SS} and \textbf{CSSinger-SS-NP}) can reduce the latency compared to the parallel inference baseline (\textbf{SiFiSinger}), while \textbf{CSSinger-FS} significantly shortens the latency during audio synthesis. Since the streaming framework processes audio chunk by chunk, \textbf{SiFiSinger} achieves the best RTF. In the CPU inference case, the RTF of the CSSinger-based models remains on par with \textbf{SiFiSinger}, while their latency metrics are significantly better than \textbf{SiFiSinger}. In the limited CPU inference case, the gap in latency between \textbf{SiFiSinger} and the chunk streaming CSSinger-related models widens even further, with \textbf{SiFiSinger} showing the lowest RTF. This indicates that the sequence length imposes a substantial computational burden in constrained environments. In both CPU scenarios, \textbf{CSSinger-FS} still achieves the lowest latency and the best RTF in the constrained case. This demonstrates that the proposed fully-streaming structure not only achieves superior or comparable synthesized audio quality to fully-parallel and semi-streaming models but also excels in latency and real-time inference efficiency. This conclusion highlights the superiority of our  method in real-time demanding or resource-constrained applications} like terminal devices.
\section{Conclusion}
In this study, {we proposed a chunk-based streaming SVS system (called CSSinger) and a fully-streaming variant (named CSSinger-FS). By incorporating strategies including Natural Padding and Causal Smooth Layer, the proposed method can effectively improve the quality of synthesized audio and significantly reduce the latency. Results demonstrate that CSSinger-FS excels in multiple objective and subjective metrics. It was shown that CSSinger-FS outperforms existing parallel and semi-streaming systems in MOS scores and objective metrics on both the Opencpop and PopCS datasets. It also exhibits a superiority in latency in both GPU and CPU hardware configurations. This is rather important for high real-time demanding applications and low computational resource scenarios.


\bibliography{aaai25}
 
\appendix
\section{Implementation Details}

Each set of experiments was trained for 500k steps on four NVIDIA V100 GPUs. The batch size for experiments on the Opencpop and Baker datasets was 16, while the batch size for experiments on the PopCS dataset was 8. It is worth noting that due to the length of each audio sample in the PopCS dataset (mostly ranging from 10 to 20 seconds), we applied random slicing to the input feature sequences before the Acoustic Model Decoder in the PopCS experiments to alleviate computational pressure.

\section{Related Work}
\subsection{Singing Voice Synthesis}
Singing Voice Synthesis (SVS) aims to generate high-quality singing voices based on music scores. Similar to Text-to-Speech (TTS) tasks, early works in singing voice synthesis employed concatenative or parametric approaches \cite{macon1997concatenation, bonada2003sample, saino2006hmm,dutoit1996use}, which often involved complex pipelines and lacked naturalness. Thanks to the rapid advancements in deep learning, many SVS systems based on deep neural networks have emerged in recent years \cite{hono2021sinsy,kim2018korean,nakamura2019singing,nishimura2016singing}. XiaoiceSing \cite{lu2020xiaoicesing} adopts an acoustic model based on FastSpeech \cite{ren2019fastspeech}, while ByteSing \cite{9362104} employs an autoregressive architecture similar to Tacotron \cite{wang2017tacotron}. VISinger \cite{zhang2022visinger} and VISinger2 \cite{zhang23e_interspeech} have proposed fully end-to-end singing voice synthesis systems based on the VITS \cite{kim2021conditional} conditional VAE architecture and adversarial training \cite{goodfellow2020generative}. SiFiSinger \cite{cui2024sifisinger} further optimizes pitch accuracy in acoustic modeling and improves the quality of synthesized singing voices, building on the foundation of VISinger2.

\subsection{Conditional VAE for Speech Synthesis}
Variational Autoencoder (VAE) \cite{kingma2013auto} is one of the most widely used deep generative models today, with numerous applications in the Speech Synthesis domain. \cite{zhang2019learning} introduced VAE into Tacotron2 \cite{shen2018natural} to learn latent representations of speaker states, enabling better control over the speaking style in synthesized speech. \cite{lee2020bidirectional} applied BVAE \cite{kingma2016improved} to TTS, proposing a non-autoregressive TTS system for generating mel-spectrograms that improves inference speed while maintaining the quality of the synthesized speech.

VITS \cite{kim2021conditional} introduced the conditional VAE framework in TTS tasks, modeling the prior distribution of latent variables based on text conditions. VITS incorporates the MAS mechanism to explicitly align latent sequences with source sequences. Unlike previous two-stage approaches, VITS is a fully end-to-end TTS system. It enhances the generative model's expressiveness using normalizing flows and variational inference, resulting in audio samples with greater naturalness and realism. The work of \cite{zhang2022visinger,zhang23e_interspeech,cui2024sifisinger} has leveraged the conditional VAE architecture from VITS for singing voice synthesis.
\begin{figure}[h]
  \hspace{-0.5cm}
  \includegraphics[width=1.1\columnwidth]{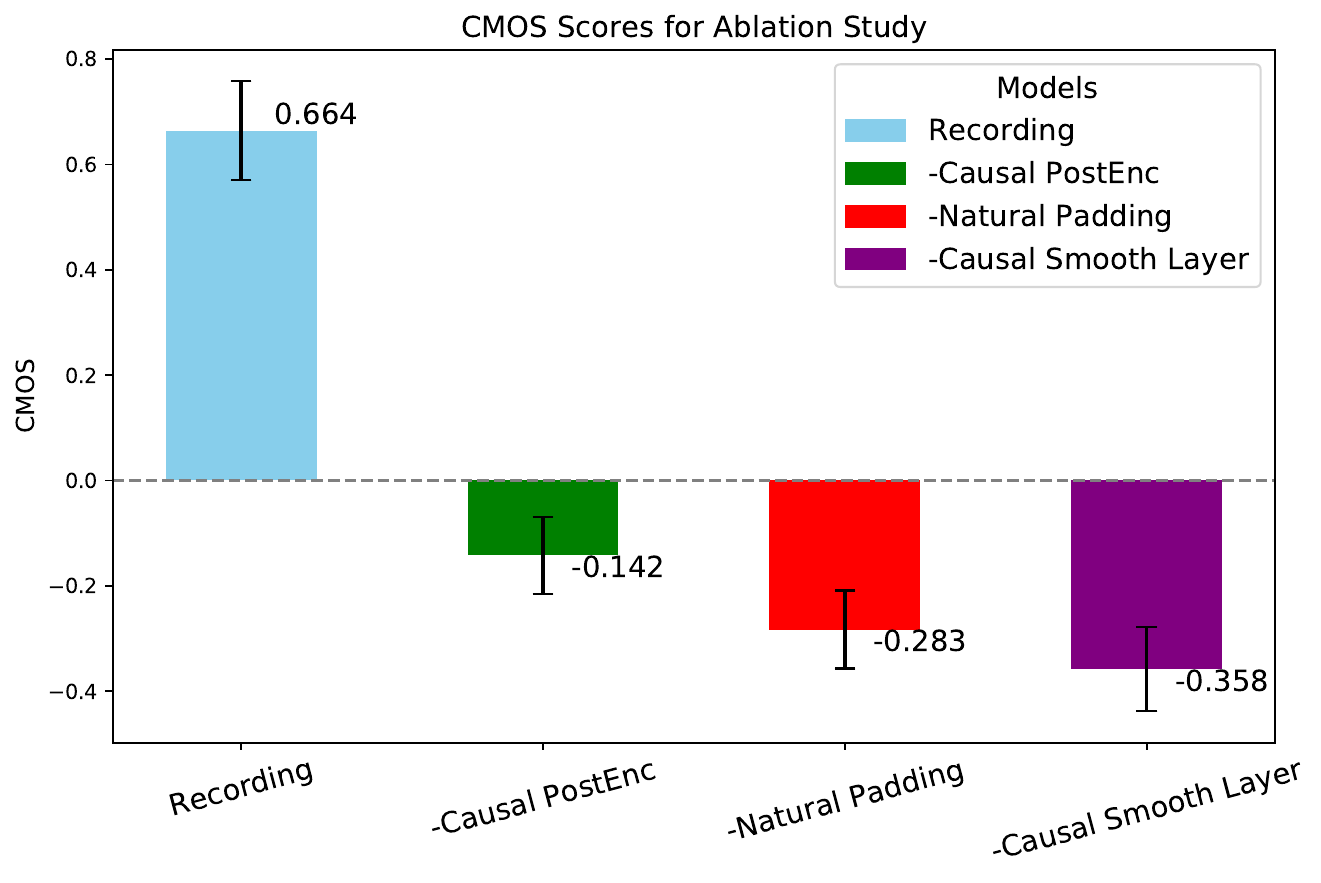}
  \caption{Bar chart of CMOS scores with 95\% confidence intervals for the ablation study}
  \label{fig:cmos}
\end{figure}
\section{Ablation Study}
To verify the effectiveness of the proposed modules, we conducted ablation experiments on the Opencpop dataset. We sequentially removed key components and conducted comparative mean opinion score (CMOS) tests and objective metric evaluations to observe their impact on model performance. The CMOS test results and objective test results are shown in Figure~\ref{fig:cmos} and Table~\ref{tab:objective ablation}, respectively.

\begin{table}[ht]
\centering
\resizebox{\columnwidth}{!}{%
\begin{tabular}{@{}llllll@{}}
\toprule
\multirow{2}{*}{\textbf{Models}} &
  \multicolumn{1}{c}{\multirow{2}{*}{\textbf{\begin{tabular}[c]{@{}c@{}}F0 \\ RMSE\( \downarrow \)\end{tabular}}}} &
  \multicolumn{1}{c}{\multirow{2}{*}{\textbf{\begin{tabular}[c]{@{}c@{}}F0 \\ corr\( \uparrow \)\end{tabular}}}} &
  \multicolumn{1}{c}{\multirow{2}{*}{\textbf{\begin{tabular}[c]{@{}c@{}}U/UV \\  Err\( \downarrow \)\end{tabular}}}} &
  \multicolumn{1}{c}{\multirow{2}{*}{\textbf{MSE}\( \downarrow \)}} &
  \multicolumn{1}{c}{\multirow{2}{*}{\textbf{MCD}\( \downarrow \)}} \\
                              & \multicolumn{1}{c}{} & \multicolumn{1}{c}{} & \multicolumn{1}{c}{} & \multicolumn{1}{c}{} & \multicolumn{1}{c}{} \\ \midrule
CSSinger-FS          & 28.101               & 0.919                & 0.107                & 1.093                & 6.715                \\
\ \ -Causal PostEnc      & 28.334               & 0.923                & 0.109                & 1.092                & 6.708                \\
\ \ \ -Natural Padding     & 30.519               & 0.874                & 0.123                & 1.097                & 6.736                \\
\ \ \ \ -Causal Smooth Layer & 37.872               & 0.826                & 0.115                & 1.109                & 6.810                \\ \bottomrule
\end{tabular}%
}
\caption{Objective metrics of the ablation study}
\label{tab:objective ablation}
\end{table}

First, we replaced the causal Posterior Encoder in CSSinger-FS with a non-causal version (\textbf{-Causal PostEnc}). The results showed a decline in both MOS scores and objective metrics. As shown in Figure~\ref{fig:melspec}(f), the model's ability to express high pitch transitions that were previously captured correctly disappeared after removing the causal Posterior Encoder, indicating that the causal Posterior Encoder helps model consistent patterns in prior and posterior distributions during training in the fully streaming framework, thus enabling the model to more accurately capture pitch variations. Next, we removed the Natural Padding(\textbf{-Natural Padding}). Consistent with previous experimental results and discussions, the quality of the generated audio decreased.

Finally, we removed the Causal Smooth Layer from the ChunkStream Decoder(\textbf{-Causal Smooth Layer}). While there was a decline in subjective metrics, it is noteworthy that the F0 RMSE showed a more significant decline compared to the previous two components. As discussed earlier and shown in Figure~\ref{fig:melspec}, the Causal Smooth Layer plays a role in smoothing the outputs of the attention layers, better handling local contextual dependencies between chunks. Despite being composed of simple one-dimensional causal convolutions, it effectively filters out discontinuities and spikes in the spectrogram, significantly enhancing the quality of the synthesized audio.

\begin{table}[ht]
\centering
\resizebox{\columnwidth}{!}{%
\begin{tabular}{@{}c|c|ccc@{}}
\toprule
\textbf{Models} & Recording & DiffSinger & CSSinger-FS & VISinger2 \\ \midrule
\textbf{MOS}    & 4.77±0.072 & 3.43±0.083  & \textbf{4.07±0.081}   & 3.83±0.091 \\ \bottomrule
\end{tabular}%
}
\caption{Horizontal comparison between various SVS systems}
\label{table:horizontal}
\end{table}
\section{Horizontal Comparison}
To demonstrate that the proposed method not only improves performance compared to the baseline but also shows superiority over other models for the same task, we conducted a horizontal comparison between the proposed model(CSSinger-FS) and two other mainstream singing synthesis systems: VISinger2\cite{zhang23e_interspeech} and DiffSinger\cite{liu2022diffsinger}. We performed a subjective MOS evaluation on the Opencpop dataset, with the results presented in Table~\ref{table:horizontal}. 

It is important to note that we reproduced VISinger2 using the official implementation, while for DiffSinger, we directly used the official checkpoint files (including the acoustic model and NSF HiFiGAN vocoder). In the original DiffSinger paper, ground truth F0 was provided for the singing voice synthesis task. However, the official code repository also offers a DiffSinger system with an integrated F0 predictor for inferring F0 values during synthesis. We followed this pipeline to generate the synthesized audio, ensuring consistency across the workflows of the comparison systems.

\end{document}